\documentclass{article}
\usepackage{spconf,amsmath,graphicx,amsthm,amssymb}
\usepackage{verbatim}
\usepackage{enumerate,mdwlist}

\usepackage{bm}
\usepackage{amsfonts}
\title{Accelerated spectral clustering using \\graph filtering of random signals}

\name{$\mbox{Nicolas Tremblay}^{\,\mbox{\scriptsize (1,2,3)}}$\thanks{This work was partly funded by the European Research Council, PLEASE project (ERC-StG-2011-277906); and the ANR-14-CE27-0001 GRAPHSIP grant.}, $\mbox{Gilles Puy}^{\,\mbox{\scriptsize (1)}}$, $\mbox{Pierre Borgnat}^{\,\mbox{\scriptsize (2)}}$, $\mbox{R\'emi Gribonval}^{\,\mbox{\scriptsize (1)}}$ and $\mbox{Pierre Vandergheynst}^{\,\mbox{\scriptsize (1,3)}}$}
\address{
$\,^{\,\mbox{\scriptsize(1)}}$  INRIA Rennes - Bretagne Atlantique, Beaulieu Campus, Rennes, France\\
$\,^{\,\mbox{\scriptsize(2)}}$  Physics Laboratory, ENS Lyon, CNRS, University of Lyon, Lyon, France\\
$\,^{\,\mbox{\scriptsize(3)}}$  Institute of Electrical Engineering, EPFL, Lausanne, Switzerland}

\begin{document}
%\ninept
%
\maketitle
\begin{abstract}
We build upon recent advances in graph signal processing 
to propose a faster spectral clustering algorithm. 
Indeed, classical spectral clustering is based on the computation of the first $k$ eigenvectors 
of the similarity matrix' Laplacian, whose computation cost, even for sparse matrices, becomes 
prohibitive for large datasets. We show that we can estimate the spectral clustering 
distance matrix without computing these eigenvectors:  
by graph filtering random signals. 
Also, we take advantage of the stochasticity of these random vectors to estimate the number of clusters $k$. 
We compare our method to classical spectral clustering on synthetic data, and 
show that it reaches equal performance while being faster by a factor at least two
for large datasets. 
\end{abstract}
\begin{keywords}
Spectral clustering, graph signal processing, graph filtering
\end{keywords}
\section{Introduction}
\label{sec:intro}
Spectral clustering has become a popular clustering algorithm, 
due to its simplicity of implementation  
and high performance for many different types of datasets~\cite{vonluxburg_StatComp2011,Jia_NCA2014}. 
Given a set of $N$ data points 
$(\bm{x}_1,\bm{x}_2,\cdots,\bm{x}_N)$, it basically transforms them  non-linearly 
into a $k$-dimensional space first, by computing a similarity matrix $\bm{W}$ from the data, and second  
by computing the $k$ first eigenvectors of its Laplacian. Calculating these eigenvectors is the 
computational bottleneck of spectral clustering: it becomes prohibitive when $N$ becomes large and/or 
when the $k$-th eigengap becomes too small~\cite{vonluxburg_StatComp2011}. Circumventing this issue is an 
active area of research~\cite{lei_AoStat2015,Kang_TKDE2014}. Another difficulty, common to 
all clustering methods, is the estimation of the usually unknown number of 
clusters $k$~\cite{Jain_1999,bendavid_chapter2006}.  

We propose a new method that jointly avoids to partially diagonalize the Laplacian and proposes 
a stability measure to estimate $k$. This method is based on the emerging field of graph signal 
processing~\cite{sandryhaila_SPMAG2014,shuman_SPMAG2013}, where the graph we consider here is defined by the 
weighted adjacency matrix $\bm{W}$. In our previous work~\cite{tremblay_TSP2014}, 
we proposed the use of the graph wavelet~\cite{tremblay_globalsip2013} or scaling function~\cite{Roux_WIFS2015}
transforms of random vectors to detect multiscale 
communities in networks. In this paper, we build upon this idea of using filtered random vectors as 
probes of the underlying graph's structure; and further prove that ideal low-pass graph filters 
have deep connections with spectral clustering. Taking advantage of the fast graph filtering 
defined in~\cite{shuman_DCOSS2011} to 
low-pass filter such random signals \emph{without computing the first $k$ eigenvectors of the graph's Laplacian},
we propose an accelerated spectral clustering method that has the collateral advantage of defining a stability 
measure to estimate the number of clusters $k$. 

In Section~\ref{sec:gsp}, we recall the 
graph signal processing notations and tools we will use in the paper, as well as 
the classical spectral clustering algorithm. In Section~\ref{sec:sp2rv}, we prove that one can filter 
random signals to estimate the spectral clustering distance matrix. In Section~\ref{sec:asc}, we 
detail our algorithm and propose a stability measure to estimate the number of clusters $k$. 
We finally show results obtained on a controled dataset in Section~\ref{sec:results}; before 
concluding in Section~\ref{sec:conclusion}.

\section{Background}
\label{sec:gsp}

Let $\mathcal{G}=(\mathcal{V},\mathcal{E},\mathbf{W})$ be an undirected weighted 
graph with $\mathcal{V}$ the set 
of $N$ nodes, $\mathcal{E}$ the set of edges, and $\mathbf{W}$ the weighted adjacency 
matrix such that $W_{ij}=W_{ji}\geq0$ is 
the weight of the edge between nodes $i$ and $j$.  %Denote $N$ the total number of nodes. 

\subsection{The graph Fourier matrix}
Consider the graph's combinatorial 
Laplacian\footnote{One could use other types of Laplacians, such as the normalized Laplacian; which   
adds a normalization step~\cite{ng_ANIPS2002} after step 2 of the spectral clustering algorithm 
(see Section~\ref{subsec:SP}) and slightly changes  our subsequent proofs.} matrix 
$\bm{L}=\mathbf{S}-\mathbf{W}$ where 
$\mathbf{S}$ is diagonal  
with $S_{ii}= s_i = \sum_{j\neq i} W_{ij}$ 
the strength of node $i$. %The normalized Laplacian matrix reads 
%$\bm{\mathcal{L}}=\mathbf{S}^{-\frac{1}{2}}\mathbf{L}\mathbf{S}^{-\frac{1}{2}}
%=\mathbf{I}_N-\mathbf{S}^{-\frac{1}{2}}\mathbf{W}\mathbf{S}^{-\frac{1}{2}}$, 
%where $\mathbf{I_N}$ is the identity matrix of size N. 
$\bm{L}$ is real symmetric, therefore diagonalizable in an orthogonal basis: its spectrum is composed 
of its set of eigenvalues $\{\lambda_l\}_{l=1\dots N}$ that we sort: 
$0=\lambda_1\leq\lambda_2\leq\lambda_3\leq\dots\leq\lambda_{N}$;  
and of $\mathbf{\bm{\chi}}$ the orthonormal matrix of its eigenvectors:
$\bm{\chi}=\left(\bm{\chi}_1|\bm{\chi}_2|\dots|\bm{\chi}_N\right)$. 
Considering only connected graphs, the multiplicity of eigenvalue $\lambda_1=0$ is one~\cite{chung_book1997}. 
By analogy to the continuous Laplacian operator whose eigenfunctions
are the classical Fourier modes and eigenvalues their 
squared frequencies, the columns of $\bm{\chi}$ are considered as the graph's 
Fourier modes, and 
$\{\sqrt{\lambda_l}\}_{l}$ as its set of associated ``frequencies''~\cite{shuman_SPMAG2013}. 
Other types of graph Fourier matrices have been proposed (e.g. \cite{sandryhaila_TSP2013}), 
but in order to exhibit the link between graph signal processing and classical spectral clustering (that 
partially diagonalizes the Laplacian matrix), the Laplacian-based Fourier matrix is more natural.

\subsection{Spectral clustering}
\label{subsec:SP}
Let us recall the method of spectral clustering~\cite{hastie_book2005,vonluxburg_StatComp2011}. 
The input is the set of data points $(\bm{x}_1,\bm{x}_2,\cdots,\bm{x}_N)$ and $k$ the number of clusters one desires. 
Follow the steps:
\begin{enumerate}
 \item Compute the pairwise similarities $s(\bm{x}_i,\bm{x}_j)$ and create a 
 similarity\footnote{See~\cite{vonluxburg_StatComp2011} for several choices of similarity measure $s$ 
 as well as several ways to create $\bm{W}$ from the $s(\bm{x}_i,\bm{x}_j)$.} graph 
 $\bm{W}$.
Compute its Laplacian $\bm{L}$. 
 \item Let $\bm{U}\in\mathbb{R}^{N\times k}$  contain $\bm{L}$'s first $k$ 
 eigenvectors: $\bm{U}=\left(\bm{\chi_1}|\bm{\chi_2}|\cdots|\bm{\chi_k}\right)$. In other words, the columns 
 of $\bm{U}$ are the first $k$ low-frequency Fourier modes of the graph.
 \item Treat each node $i$ as a point in $\mathbb{R}^k$ by defining its feature vector $\bm{f}_i\in\mathbb{R}^k$ 
 as the $i$-th row of $\bm{U}$:
 \begin{equation}
  \bm{f}_i=\bm{U}^\top\bm{\delta}_i,
 \end{equation}
where $\delta_i(j)=1$ if $j=i$ and $0$ elsewhere.
 \item Run $k$-means (or any clustering algorithm) with the Euclidean distance 
% \begin{equation}
  $D_{ij}=||\bm{f}_i-\bm{f}_j||$ 
 %\end{equation}
to obtain 
 $k$ clusters.
\end{enumerate}

\subsection{Graph filtering}
\label{subsec:filt}
The graph Fourier transform $\bm{\hat{x}}$ of a signal $\bm{x}$ defined 
on the nodes of the graph reads: $\bm{\hat{x}}=\bm{\chi}^\top \bm{x}$. 
Given a continuous filter function $h$ defined on $[0,\lambda_N]$, its associated graph filter 
operator $\bm{H}\in\mathbb{R}^{N\times N}$ is defined as:
 $\bm{H}=\bm{\chi}\widehat{\bm{H}}\bm{\chi}^\top$,
where $\widehat{\bm{H}}=\mbox{diag}(h(\lambda_1),h(\lambda_2),\cdots,h(\lambda_N))$. We write $\bm{Hx}$ 
the signal $\bm{x}$ filtered by $h$. In the following, we will consider ideal low-pass filters, 
generically noted $h_{\lambda_c}$, such that:
\begin{equation}
\label{eq:ideal}
  h_{\lambda_c}(\lambda)=1 \mbox{ if } \lambda\leq \lambda_c ~~~~\mbox{    and     }~~~~
  h_{\lambda_c}(\lambda)=0 \mbox{ if not},
\end{equation}
and $\bm{H}_{\lambda_c}$ its associated graph filter operator. 

\subsection{Fast graph filtering}
\label{subsec:fast_filt}
In order to filter a signal by $h$ without diagonalizing the Laplacian matrix, one may rely~\cite{shuman_DCOSS2011} on a 
polynomial approximation of order $m$ 
of $h$ on $[0,\lambda_N]$:
%\begin{equation}
 $\exists\{\alpha_l\}_{l\in[0,m]}\mbox{ s.t. } h(\lambda)\simeq\sum_{l=0}^m\alpha_l\lambda^l$.
 %\end{equation}
This enables us to approximate $\bm{Hx}$:
\begin{equation}
\bm{Hx}=\bm{\chi}\widehat{\bm{H}}\bm{\chi}^\top\bm{x}\simeq\sum_{l=0}^m\alpha_l\bm{L}^l\bm{x}=\mathcal{F}^m_h\bm{x},
\end{equation}
where we note that the approximation only requires matrix-vector multiplication and effectively 
bypasses the diagonalisation of the Laplacian; and where we introduce the notation $\mathcal{F}^m_h$ 
to design the fast filtering operator of order $m$ associated to $h$. 
This fast filtering method has a total complexity of $O(m(|\mathcal{E}|+N))$~\cite{shuman_DCOSS2011}, which 
in the case of sparse graph where $|\mathcal{E}|\sim N$, ends up being $O(mN)$; compared to 
the $O(N^3)$ complexity needed to diagonalize the Laplacian matrix. \\

\noindent{\textbf{Note on the choice of $m$.}} The lower we choose $m$, the faster 
is the computation, but the less precise is the polynomial approximation. Fixing a maximal error of approximation
to $\delta$, we note $m^*$ the minimal value of $m\in\mathbb{N}^*$ such that: %for which the approximation error is bounded by $\delta$, i.e.:
\begin{equation}
\sup\limits_{\lambda\in[0,\lambda_N]} \left|h(\lambda)-\sum_{l=0}^{m}\alpha_l\bm{L}^l\bm{x}\right|\leq\delta.
\end{equation}
In this study, we only consider ideal low-pass $h_{\lambda_c}$ defined on $[O,\lambda_N]$. 
Notice that $h_{\lambda_c/\lambda_N}=h_{\lambda_c}(\lambda\times\lambda_N)$ is always defined on $[0,1]$ and 
does not depend on any graph anymore. One can therefore \textit{beforehand} tabulate $m^*$ as 
a function of its sole parameter $\lambda_c/{\lambda_N}$ (and $\delta$). Then, given a real filter 
$h_{\lambda_c}$ defined on $[O,\lambda_N]$ to approximate, one only needs to refer to this table and 
choose $m=m^*(\lambda_c/\lambda_N;\delta)$. In the following, we fix $\delta=0.1$.

\section{Accelerating spectral clustering}
\label{sec:sp2rv}

%One thereby ends up with a clustering for each value of $k$ between 2 and $K$. Then, with arguments 
%based on information or stability measures (or others), the user decides which clustering to keep. 

\subsection{Filtering random signals to estimate $D_{ij}$}

We show that we can estimate the distance $D_{ij}=||\bm{f}_i-\bm{f}_j||$ 
by filtering  a few random signals with the ideal low-pass $h_{\lambda_k}$ 
(as defined in Eq.~\eqref{eq:ideal} with $\lambda_c=\lambda_k$). 
First of all, consider the matrix operator $\bm{H}_{\lambda_k}$ associated to $h_{\lambda_k}$. One may write:
\begin{equation}
  \bm{H}_{\lambda_k}=\bm{\chi}\begin{pmatrix}
\bm{I}_k & \bm{0} \\
\bm{0} & \bm{0}
\end{pmatrix}\bm{\chi}^\top=\bm{U}\bm{U}^\top,
\end{equation}
where $\bm{I}_k$ is the identity of size $k$, and $\bm{0}$ null block matrices. 

Second, consider the matrix $\bm{R}=\left(\bm{r}_1|\bm{r}_2|\cdots|\bm{r}_\eta\right)\in\mathbb{R}^{N\times\eta}$ 
consisting of $\eta$ 
random signals $\bm{r}_i$, whose components are independent random Gaussian variables of zero mean and  
variance $1/\eta$. 
Consider its filtered version $\bm{H}_{\lambda_k}\bm{R}\in\mathbb{R}^{N\times\eta}$, and 
define node $i$'s feature vector $\tilde{\bm{f}}_i\in\mathbb{R}^\eta$ as the $i$-th line of this filtered matrix: 
\begin{equation}
 \tilde{\bm{f}}_i=(\bm{H}_{\lambda_k}\bm{R})^\top\bm{\delta}_i.
\end{equation}

\noindent\textbf{Proposition:}
\textit{Let $\epsilon,\beta>0$ be given. If $\eta$ is larger than:
\begin{equation}
 \eta_0=\frac{4+2\beta}{\epsilon^2/2-\epsilon^3/3}\log{N},
\end{equation}
then with probability at least
  $1-N^{-\beta}$,
 we have: $\forall(i,j)\in[1,N]^2$
\begin{equation}
\label{eq:bounding}
(1-\epsilon)||\bm{f}_i-\bm{f}_j||^2\leq ||\tilde{\bm{f}}_i-\tilde{\bm{f}}_j||^2\leq (1+\epsilon)||\bm{f}_i-\bm{f}_j||^2.
\end{equation}
}

%$\displaystyle\lim_{\eta\to\infty} \mbox{\texttt{Euc}}(\widetilde{\bm{f}_i},\widetilde{\bm{f}_j})=D_{ij}^{SP}$.

\begin{proof}
We rewrite $||\tilde{\bm{f}}_i-\tilde{\bm{f}}_j||^2$ in a form that will let us apply the 
Johnson-Lindenstrauss lemma of norm conservation: 
\begin{equation}
\label{eq:norm}
\begin{aligned}
 ||\tilde{\bm{f}}_i-\tilde{\bm{f}}_j||^2&=||\bm{R}^\top\bm{H}_{\lambda_k}^\top(\bm{\delta}_i-\bm{\delta}_j)||^2\\
 &=||\bm{R}^\top\bm{U}\bm{U}^\top(\bm{\delta}_i-\bm{\delta}_j)||^2\\
 &=||\bm{R}^\top\bm{U}(\bm{f}_i-\bm{f}_j)||^2.
\end{aligned}
\end{equation}
As i)~the columns of $\bm{U}$ are normalized to 1 and orthogonal to each other, 
and ii)~$\bm{R}$ is a random Gaussian matrix with mean zero and variance $1/\eta$, 
then $\bm{R}'=\bm{R}^\top\bm{U}$ is also Gaussian with same mean and variance; and 
 Equation~\eqref{eq:norm} reads:
 \begin{equation}
\label{eq:for_jl}
 ||\tilde{\bm{f}}_i-\tilde{\bm{f}}_j||^2=||\bm{R}'(\bm{f}_i-\bm{f}_j)||^2.
\end{equation}
This enables us to apply Theorem 1.1 
of~\cite{Achlioptas_JCSS2003} (an instance of the Johnson-Lindenstrauss lemma) and finish the proof.
\end{proof}
\begin{comment}
As $\bm{R}$ is a random Gaussian matrix with mean zero and variance $1/\eta$, then 
$\bm{R}'=\bm{R}^\top\bm{U}$ is also a random Gaussian matrix with same mean and variance. 
Indeed, any element of $\bm{R}'$ reads: $R_{ij}'=\sum_l R_{li}U_{lj}$. 
Therefore its expected value reads:
\begin{equation}
 \mathbb{E}(R_{ij}')=\sum_l \mathbb{E}(R_{li})U_{lj}=0,
\end{equation}
and its variance:
\begin{equation}
\begin{aligned}
 \mathbb{E}(R_{ij}'^2)&=\sum_{l\neq l'} U_{lj}U_{l'j}\mathbb{E}(R_{li}R_{l'i})+\sum_{l} U_{lj}^2\mathbb{E}(R_{li}^2)\\
 &=0+\frac{1}{\eta}\sum_{l} U_{lj}^2=\frac{1}{\eta},
\end{aligned}
\end{equation}
as the columns of $\bm{U}$ are normed to 1.  
\end{comment}

\noindent\textbf{Consequence:} Setting $\beta$ to 1, and therefore the failure probability of  
Equation~\eqref{eq:bounding} to $1/N$, we only need to filter $\eta\gtrsim\frac{12}{\epsilon^2}\log{N}$ random signals
to estimate (up to an error $\epsilon$) the spectral clustering distance matrix. How this error 
$\epsilon$ on 
the distance estimation theoretically affects the performance of the spectral clustering algorithm 
is still, to our knowledge, an open question. We observe experimentally (see Sec.~\ref{sec:results}) that using a number 
$\eta\gtrsim k$ (i.e. allowing a relatively high error 
$\epsilon^2\simeq12\frac{\log{N}}{k}$) is usually enough for satisfying performance.

\subsection{Fast filtering of random signals}
\label{subsec:ffrv}
In practice, we do not exactly filter these random signals by $h_{\lambda_k}$ as the 
computation of $\bm{H}_{\lambda_k}$ 
requires the diagonalisation 
of the Laplacian, which is precisely what we are trying to avoid. 
Instead, we take advantage of the fast 
filtering scheme recalled in Section~\ref{subsec:fast_filt}. Still, one question remains: 
the fast filtering is based on the polynomial approximation of $h_{\lambda_k}$, which is itself 
parametrized by $\lambda_k$. Unless we compute the first $k$ eigenvectors of $\bm{L}$, thereby loosing 
our efficiency edge on other methods, we cannot know \textit{exactly} the value of $\lambda_k$. 

To circumvent this, we estimate the spectrum's cumulative density function as in Section VB 
of~\cite{shuman_TSP2015} (the \textit{cdf} of a graph's spectrum is roughly linear only if the graph shows  
topological regularity; if not, it is not trivial). Given a graph Laplacian $\bm{L}$, 
and a ``cutting frequency'' $\lambda_c$, 
the algorithm estimates the number of 
eigenvalues inferior to $\lambda_c$. Using a dichotomous procedure on $[0,\lambda_N]$, we  
obtain an estimate of $\lambda_k$, noted $\tilde{\lambda}_k$. 

\begin{comment}
\subsection{Stability measure to estimate $\lambda_k$}
We will consider $s$ different values of the ``cutting frequency'':
$$\lambda_c\in\Lambda=\{\lambda_c^1,\lambda_c^2,\cdots,\lambda_c^s\},$$ 
and compute, for each of these $s$ values, the clustering algorithm $J$ times using $J$ different realisations 
of the $\eta$ random vectors, to obtain $J$ different clusterings $\{\mathcal{C}^j_{\lambda_c}\}_{j\in[1,J]}$. 
For each value of $\lambda_c$, we then define a notion of stability 
as the mean of the Adjusted Rand Index similarity~\cite{Hubert1985} (noted \texttt{ari}) 
 between all pairs of clusterings:
\begin{equation}
\label{eq:stab}
 \gamma(\lambda_c)=\frac{2}{J(J-1)}\displaystyle\sum_{(i,j)\in[1,J]^2,i\neq j}\mbox{\texttt{ari}}(\mathcal{C}^i_{\lambda_c},\mathcal{C}^j_{\lambda_c})
\end{equation}
We then look at the global maximum of $\gamma$ to obtain our estimation of $\lambda_k$, and more importantly, 
the clustering we are looking for. 

The last question we need to address is how to create the sampling set $\Lambda$? 
Quite naively, we propose $\Lambda$ to be a regular sampling between 0 (excluded) and 1. 
Indeed, eigenvectors corresponding to eigenvalues larger than 1 are not linked to the cluster structure of the 
graph, but rather to its noise. 
\end{comment}

\section{Accelerated spectral clustering}
\label{sec:asc}
\subsection{Algorithm}
\label{sec:algo}
Consider a set of data points $(\bm{x_1},\bm{x_2},\cdots,\bm{x_N})$ and $k$ the number of desired clusters. 
The first step of the algorithm does not change as compared to Section~\ref{subsec:SP}: 
compute the pairwise similarities $s_{ij}$, create a similarity graph $\bm{W}$,  
and compute its Laplacian $\bm{L}$. 
Then: 
\begin{enumerate}
 \item Estimate~\cite{trefethen1997numerical} $\bm{L}$'s largest eigenvalue $\lambda_N$ (necessary for steps 2 and 3).
 \item Estimate $\bm{L}$'s $k$-th eigenvalue $\tilde{\lambda}_k$ as in 
 Sec.~\ref{subsec:ffrv}.
 \item Set $m=m^*(\tilde{\lambda}_k/\lambda_N;\delta)$ as in Sec.~\ref{subsec:fast_filt}; 
 compute the polynomial approximation of order $m$ of the ideal low-pass filter 
 $h_{\tilde{\lambda}_k}$ to obtain the fast filtering operator 
 $\mathcal{F}^m_{\tilde{\lambda}_k}$.
 \item Generate $\eta$ random Gaussian signals of mean $0$ and variance $1/\eta$: 
 $\bm{R}=\left(\bm{r}_1|\bm{r}_2|\cdots|\bm{r}_{\eta}\right)\in\mathbb{R}^{N\times\eta}$. 
 \item Filter these signals with $\mathcal{F}^m_{\tilde{\lambda}_k}$% (as in Sec.~\ref{subsec:fast_filt}) 
 and define, for each node $i$, its feature vector $\tilde{\bm{f}}_i\in\mathbb{R}^\eta$:
 \begin{equation}
  \tilde{\bm{f}}_i^\top=\bm{\delta}_i^\top\mathcal{F}^m_{\tilde{\lambda}_k}\bm{R}.
 \end{equation}
 \item Run $k$-means (or any clustering algorithm) with the Euclidean distance 
 %\begin{equation}
  $\tilde{D}_{ij}=||\tilde{\bm{f}}_i-\tilde{\bm{f}}_j||$ 
 %\end{equation}
to obtain 
 $k$ clusters.
\end{enumerate}

\subsection{Complexity considerations}
\label{subsec:complexity}
We compare the time complexity of this algorithm and the classical spectral clustering algorithm. 
Let us separate our algorithm (resp. the classical algorithm) into two parts: the spectral estimation part 
consisting of steps 1 and 2 (resp. step 2) and the clustering part consisting in steps 4 to 6 (resp. steps 3 and 4). 
According to~\cite{Xu_2005}, $k$-means has complexity $O(\eta N)$ where $\eta$ is the dimension of 
the $N$ considered vectors. Considering the discussion of Section~\ref{subsec:fast_filt}, 
the time complexity of the clustering part is thus $O(\eta (m+1)N)$ (resp. $O(kN)$). The time complexity of the 
spectral estimation part is difficult to estimate, as it depends~\cite{vonluxburg_StatComp2011} on 
the graph-dependent eigengap $\lambda_{k+1}-\lambda_{k}$. In both algorithms, the larger is this eigengap, 
the faster is the convergence. Nevertheless, empirical observations 
show that, even though the clustering part of the classical algorithm is faster than ours; 
our algorithm makes up to that difference by computing even faster (especially for large $N$)
its spectral estimation part than the classical one.

\subsection{Estimate the number of clusters $k$}
In real data, the number of clusters $k$ to find is usually unknown. Instead, 
one has access to a (possibly large) interval of values $[k_{min},k_{max}]$.  
Notions of  stability to estimate $k$ have been used in various 
contexts~\cite{bendavid_chapter2006,benhur_biocomp2002}; and we propose here a new one that naturally comes from the random vectors' stochasticity. 
For all $k\in[k_{min},k_{max}]$, 
we perform our algorithm $J$ times using $J$ different realisations 
of the $\eta$ random signals, to obtain $J$ different clusterings $\{\mathcal{C}^j_{k}\}_{j\in[1,J]}$. 
We define a  stability measure 
as the mean of the Adjusted Rand Index similarity~\cite{hubert_Jclassif1985}   
 between all pairs of clusterings~\cite{tremblay_TSP2014,tremblay_globalsip2013}:
\begin{equation}
\label{eq:stab}
 \gamma(k)=\frac{2}{J(J-1)}\displaystyle\sum_{(i,j)\in[1,J]^2,i\neq j}\mbox{\texttt{ari}}(\mathcal{C}^i_{k},\mathcal{C}^j_{k}).
\end{equation}
Denote $k^*$ the value of $k$ for which $\gamma$ reaches its global maximum: we consider it as the relevant number 
of clusters.

%Moreover, fast filtering $\eta$ signals has a complexity of $O(\eta m(|\mathcal{E}|+N))$. 

\section{Results}
\label{sec:results}
\begin{figure}[bt]
\centering
\hfill \includegraphics[width=0.49\linewidth]{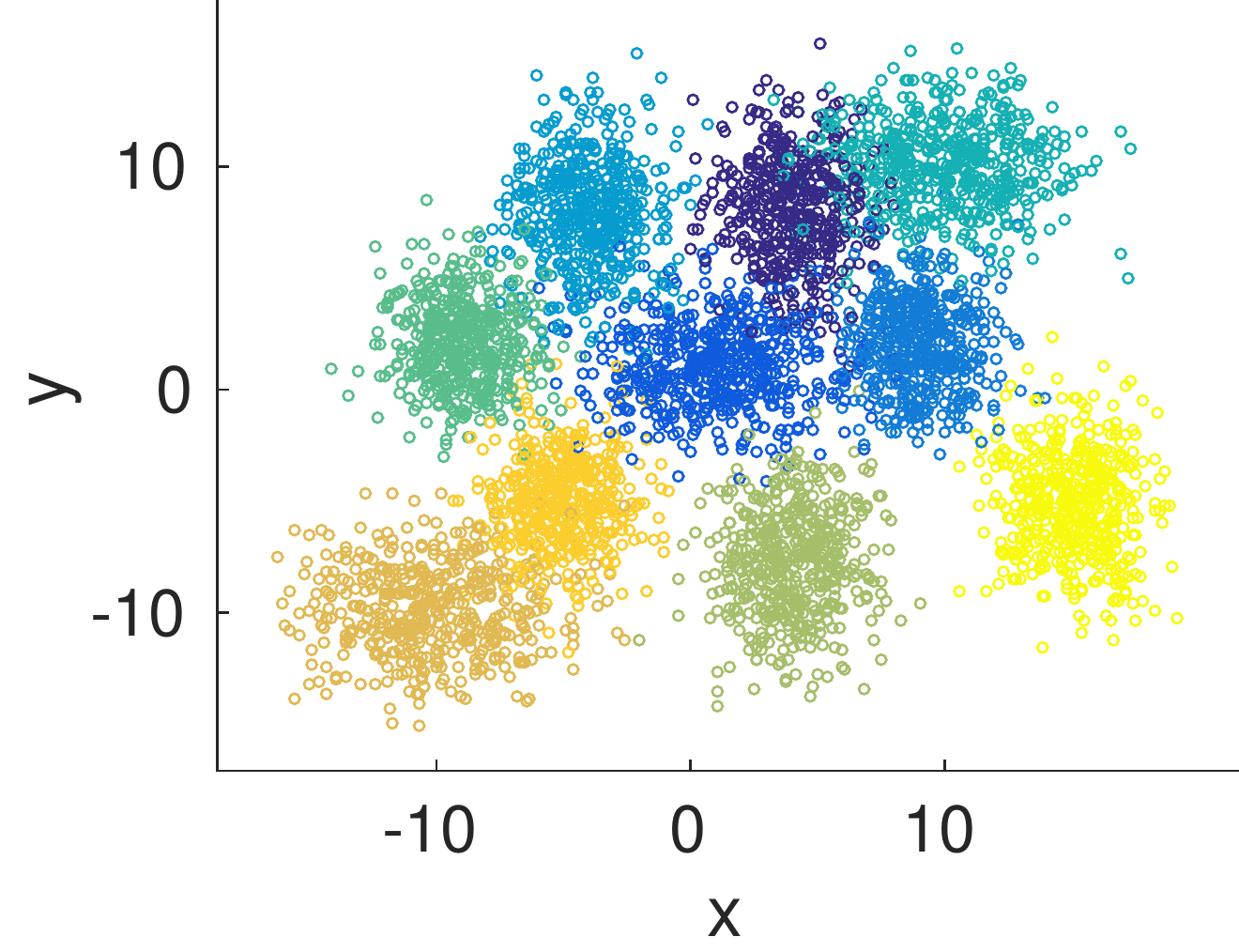}\hfill
\includegraphics[width=0.50\linewidth]{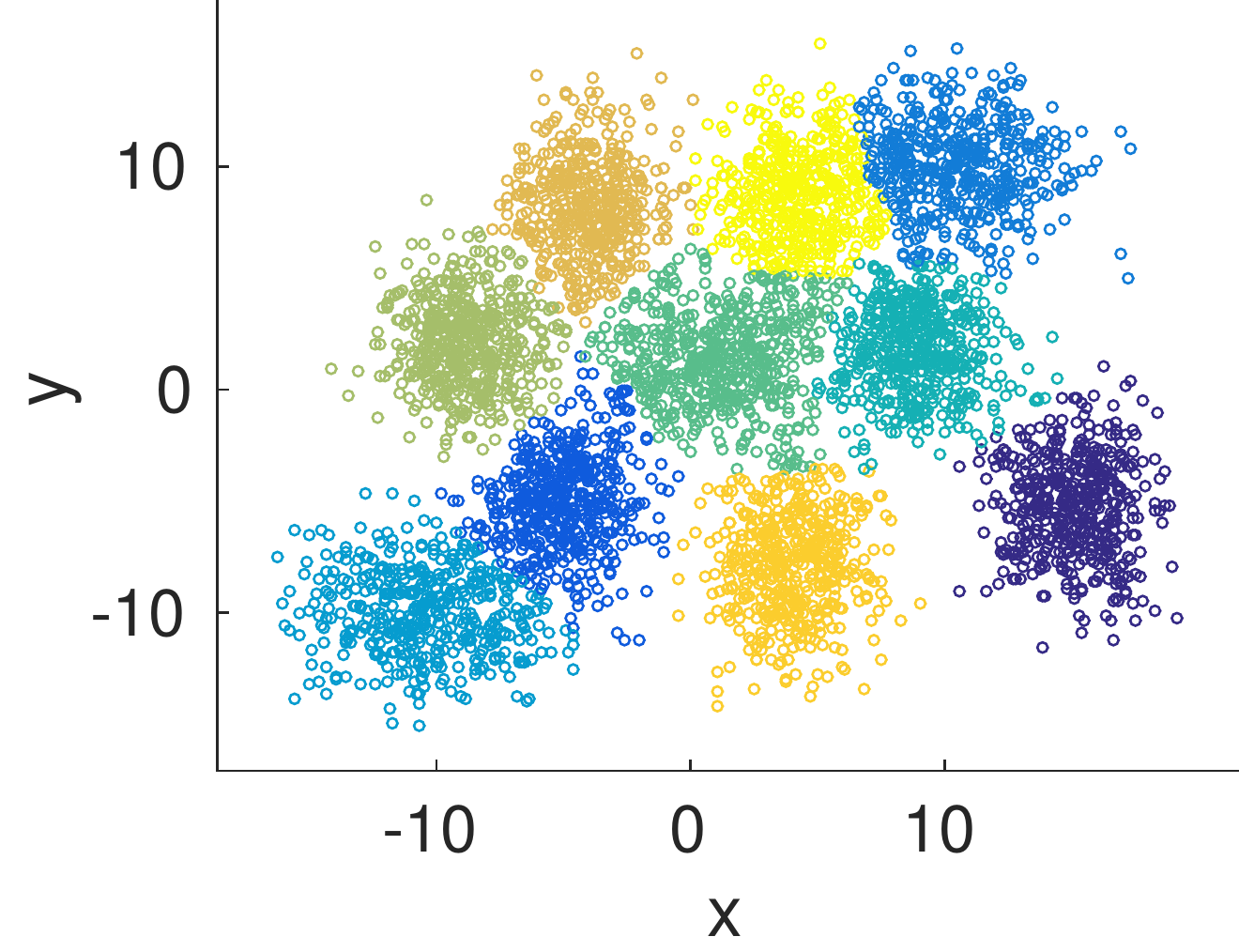}\hfill
\caption{(left) One realisation of Gaussian synthetic data with $N=5000$ and $k=10$. Colours indicate 
the label of each point. (right) Recovered labeling with our method 
and $\eta=2k$ (recovery performance $p=0.85$). A similar result is obtained using  
the original spectral clustering algorithm.
}
\label{fig:example_gaussians}
\end{figure}

We consider a sum of $k=10$ two-dimensional Gaussians with different means and variances, and from this 
distribution, draw a set of $N$ points in 
$\mathbb{R}^2$. Each point $i$ has a label $l(i)$ indicating from which Gaussian it was drawn. 
Fig.~\ref{fig:example_gaussians} (left) shows a realisation of such distribution with $N=5000$ where  
colours indicate the label of each node. The goal here is to recover the original labeling $l$. To measure how close the output $l_m$ of a given method recovers the original labeling, 
we compute the Adjusted Rand Index $p(l_m)=\texttt{ari}(l,l_m)\in[-1,1]$: the closer is the 
performance $p$ to 1, the closer is $l_m$ to $l$.

From these N points, we create a similarity matrix by building a $K$ nearest neighbours graph with $K~\sim\log{N}$ 
as suggested in~\cite{vonluxburg_StatComp2011} as a classical way of generating sparse similarity graphs. 
Other wiring possibilities exist to create such a graph (see~\cite{vonluxburg_StatComp2011}) but we only show 
this particular one as the choice of construction does not affect our results (but still considering graphs 
with same sparsity). 

\begin{figure}[bt]
\centering
\includegraphics[width=0.95\linewidth]{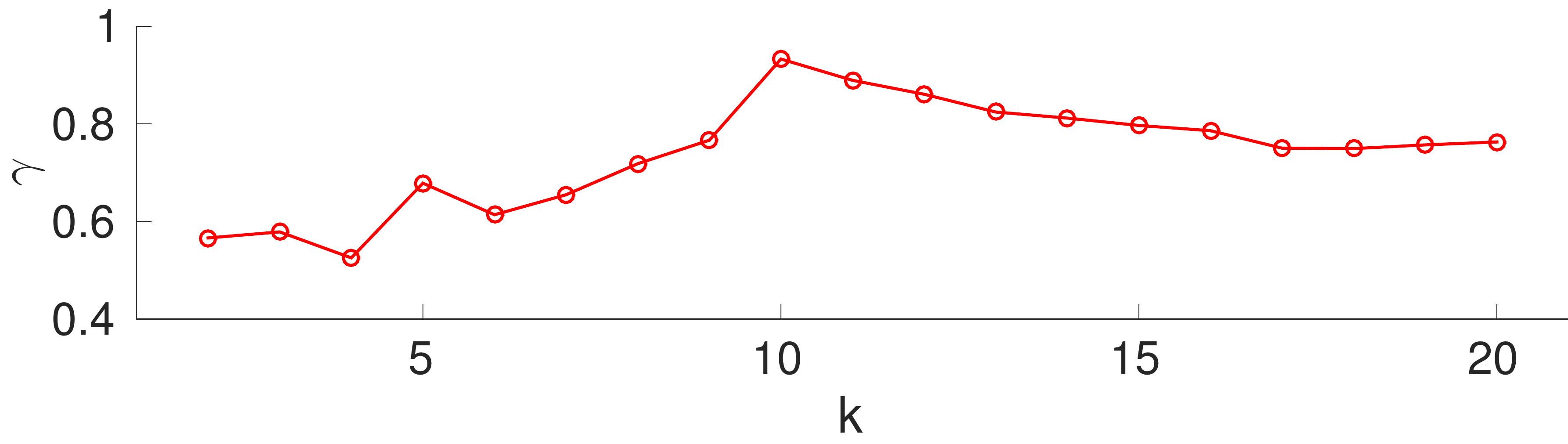}
\caption{Stability measure $\gamma$ versus the number of clusters $k$ for the Gaussian dataset 
illustrated in Fig.~\ref{fig:example_gaussians}: its maximum correctly detects $k=10$. $N=5000, \eta=2k$ and 
$J=20$.
}
\label{fig:stability}
\end{figure} 

We use Matlab's  $k$-means function (with 20 replicates) and 
the GSP toolbox~\cite{perraudin2014gspbox} for steps 2 and 3 of our algorithm. 

First, we remark in this dataset that values of $\lambda_{k=10}/\lambda_N$ are small 
(of the order of $10^{-4}$), which necessitates a large $m\sim200$ for a correct 
approximation of the ideal low-pass on $[0,\lambda_N]$ (see Section~\ref{subsec:fast_filt}). 

Let us now illustrate in Fig.~\ref{fig:stability} the stability measure $\gamma$ obtained 
using our method with $N=5000$, $\eta=2k$ and $J=20$: the global maximum correctly detects $k=10$ and 
Fig.~\ref{fig:example_gaussians} (right) shows one recovered labeling for $k=10$. 
%The second highest local maximum is for $k=5$, and corresponds to a clustering where original clusters are 
%associated two by two (not shown).

\begin{figure}[bt]
\centering
\includegraphics[width=0.95\linewidth]{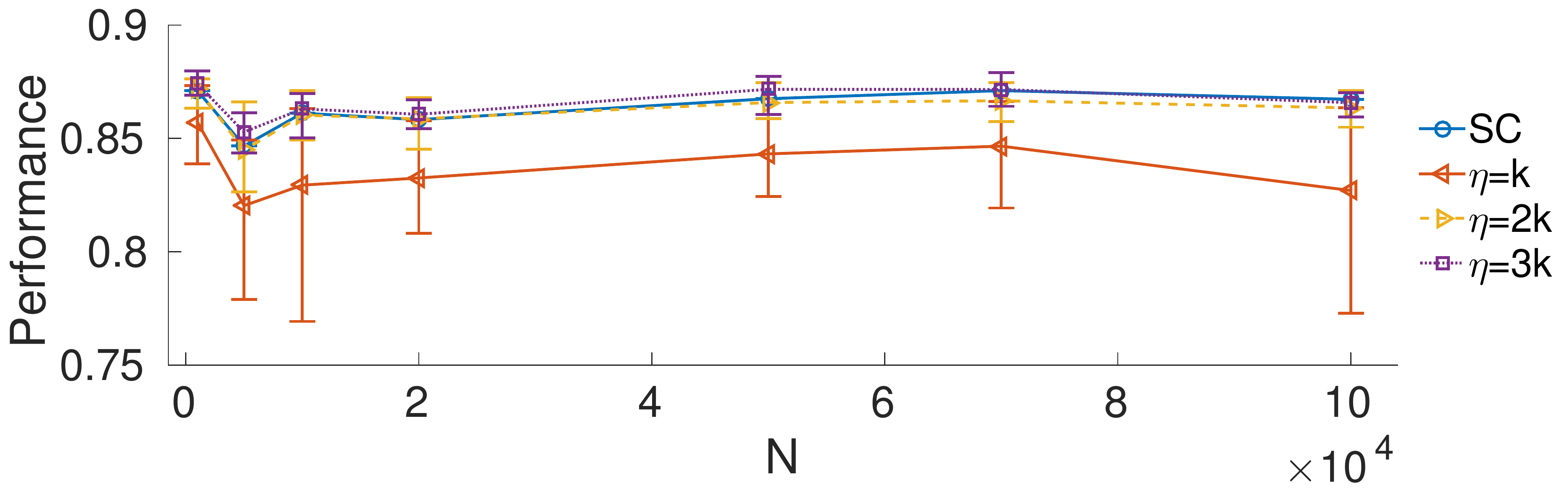}\\
\includegraphics[width=0.95\linewidth]{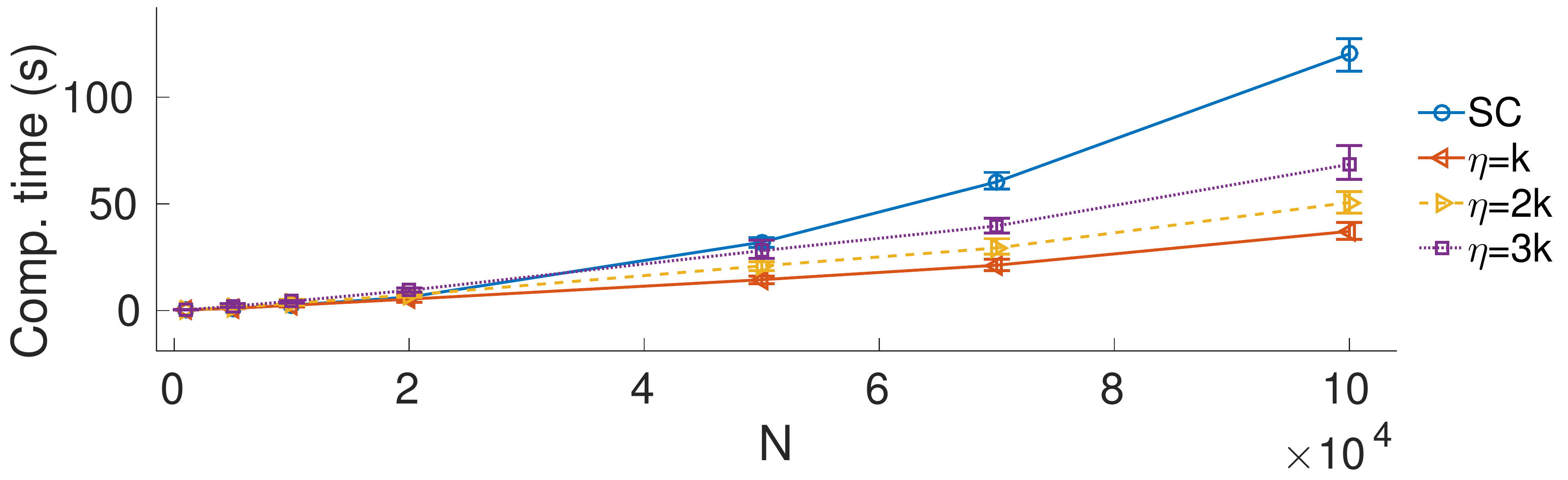}
\caption{Average (and $10$ and $90\%$ quantiles) performance (top) and computation time 
(bottom) over 20 realisations of the Gaussian dataset illustrated in 
Fig.~\ref{fig:example_gaussians} for 
the classical spectral clustering algorithm (SC), and our 
proposition for $\eta=k, 2k$ and $3k$. %Performance is in log-lin scale, and computation times in log-log scale.
%The bottom plot shows how the computation time changes if one looks 
%for $k=11$ clusters instead of $k=10$ on the same dataset.
}
\label{fig:comparisons}
\end{figure}

We compare our method vs. classical spectral clustering 
in Figure~\ref{fig:comparisons} in terms of 
performance and time of computation. We note that for $\eta=2k$ and $\eta=3k$, our method performs as well as 
the classical algorithm; while for $\eta=k$, on the other hand, 
the number of random signals becomes insufficient as we observe the recovery starting to fail. 
Up to $N\simeq 4.10^4$, the computation time is slightly faster with the classical algorithm. But as $N$ increases, 
the classical algorithm's computing time increases significantly faster than our proposition's: 
for $N=10^5$ for instance, 
 computation time is 2 (resp. 2.5, 3) times faster when one uses $\eta=3k$ (resp. $2k$, $k$) random signals 
than the classical algorithm. 

%Moreover, the bottom plot of Fig.~\ref{fig:comparisons} illustrates how the computation time changes if one looks 
%for $k=11$ clusters instead of $k=10$ on the same datasets: our method's times barely change, whereas 
%the classical spectral clustering's time doubles. As noted earlier, this is because the larger is the 
%eigengap $\lambda_{k+1}-\lambda{k}$, the faster it is to compute the first 
%$k$ eigenvectors. Our method is not sensitive to eigengaps, which is a great advantage especially in a 
%realistic context where one needs to try different values of $k$, with potentially very small eigengaps.

\section{Conclusion}
\label{sec:conclusion}
We propose a new method that paves the way to alternative spectral clustering methods bypassing the 
usual computational bottleneck of extracting the Laplacian's first $k$ eigenvectors. We take advantage 
of the fast graph low-pass graph filtering of a few random vectors to estimate the spectral clustering 
distance. The use of random vectors makes our algorithm stochastic, which in turn enables us to define a 
stability measure $\gamma$ for any $k$: scales of interest maximize $\gamma$. Results on synthetic data show that 
our method is scalable and for $\eta\gtrsim k$, one has the same performance as with the classical spectral algorithm  
while reducing the time complexity by a few factors. 

We prooved that the error 
on the estimation of the distance $D_{ij}$ is well controlled, but the question of how such an error  
propagates on the estimation of the clusters themselves is open. Moreover, the impact of the error $\delta$ 
of the polynomial approximation on the rest of the algorithm is still largely unknown and matter of future work.

%\begin{figure}[bt]
%\centering
%\includegraphics[width=0.9\linewidth]{fig/approx_filters.pdf}
%\caption{Beautiful caption.
%}
%\label{fig:approx_filters}
%\end{figure} 

% References should be produced using the bibtex program from suitable
% BiBTeX files (here: strings, refs, manuals). The IEEEbib.bst bibliography
% style file from IEEE produces unsorted bibliography list.
% -------------------------------------------------------------------------
\newpage
\bibliographystyle{IEEEbib}
\bibliography{new_biblio.bib}

\end{document}